\documentclass[epj,nopacs]{svjour}
\usepackage{graphics}
\usepackage{graphicx}
\usepackage{float}
\usepackage{bm}
\usepackage{verbatim}
\usepackage{tensor}
\usepackage{undertilde}
\usepackage{cite}
\usepackage{latexsym}
\usepackage{amsmath}
\usepackage{epstopdf}
\usepackage{widetext}
\usepackage{cite}
\newcommand{\be}{\begin{equation}}
\newcommand{\ee}{\end{equation}}

\begin{document}
\title{Effects of a string cloud on the criticality and efficiency of AdS black holes as heat engines}
\author{J. P. Morais Gra\c{c}a\inst{1} \and Iarley P. Lobo\inst{1} \and Valdir B. Bezerra\inst{1} \and H. Moradpour\inst{2}%
}                     % Do not remove
\institute{Departamento de F\'{i}sica, Universidade Federal da Para\'{i}ba, Caixa Postal 5008, CEP 58051-970, Jo\~{a}o Pessoa, PB, Brazil, \email{jpmorais@gmail.com}, \email{iarley\_lobo@fisica.ufpb.br}, \email{valdir@fisica.ufpb.br} \and Research Institute for Astronomy and Astrophysics of Maragha (RIAAM), P.O. Box 55134-441, Maragha, Iran, \email{h.moradpour@riaam.ac.ir}}
\abstract{
We study the black hole thermodynamics in the presence of a string cloud matter distribution, considering a work term due to a variable cosmological ``constant" in arbitrary dimensions. Then, we explore the criticality of the system and the behavior of the black hole as a heat engine in the context of general relativity and metric $f(R)$ gravity.
%\PACS{
   %   {PACS-key}{discribing text of that key}   \and
      %{PACS-key}{discribing text of that key}
     %} % end of PACS codes
} %end of abstract
\mail{jpmorais@gmail.com}
\titlerunning{Effects of a string cloud on the criticality and efficiency of AdS black holes}
\authorrunning{J.P.M.G., I.P.L, V.B.B., H.M.}
\maketitle

\section{Introduction}
Since the proof that the area of a black hole never classically decreases, it has been proposed that it can be interpreted as the entropy of a thermodynamic system \cite{Bekenstein:1973ur,Bardeen:1973gs}. In this framework, one can construct a dictionary between gravitational and thermodynamic concepts, e.g, through the relation between mass and the internal energy, surface gravity and the temperature, and the area of the event horizon and the entropy. In the case of the Schwarzschild black hole one cannot naively introduce a work term, but for a more complex black hole such as the Kerr-Newman, work terms can be easily introduced based on the charge and angular momentum of the black hole. One feature of this classic black hole thermodynamics is the lack of the $P dV$ mechanical work term. This missing term has been fully reintroduced only in the previous decade, by considering the cosmological constant as a state parameter, and not as a true constant\cite{Kastor:2009wy}.

In this extended framework, the pressure of the system is related to the cosmological constant, and the gravitational mass is no longer identified with the internal energy, but with the enthalpy. In this scenario, there is a possibility of the system to absorb heat and to perform mechanical work, and thus a next natural step would be to study the black hole itself as a heat engine. This idea has been put forward by C. Johnson for the case of the charged Schwarzschild-Anti de Sitter (AdS) metric \cite{Johnson:2014yja}, and since then several authors have studied other kinds of black holes both in general relativity and in some modified theories of gravity \cite{
Jafarzade:2017sur,Setare:2015yra,Johnson:2015ekr,Johnson:2015fva,Bhamidipati:2016gel,Zhang:2016wek,Chakraborty:2016ssb,Hennigar:2017apu,Liu:2017baz,Xu:2017ahm,Hendi:2017bys,Wei:2017vqs,Chakraborty:2017weq,Zhang:2018vqs,Rosso:2018acz,Mo:2018hav,EslamPanah:2018ums,Belhaj:2015hha}. The main purpose of this paper is to give our contribution to the subject, studying how the inclusion of a cosmological cloud of strings, in the background of an asymptotically anti-de Sitter black hole, would affect the correlated thermodynamic system and its configuration as a heat engine. 

Another interesting result concerning the introduction of a mechanical work term is the possibility to study phase transitions on the $P-V$ plane. The first phase transition on the history of black hole thermodynamics has been discovered by Hawking and Page in their seminal paper on the thermodynamics of AdS black holes\cite{Hawking:1982dh}. With the realization of Maldacena \cite {Maldacena:1997re} and others that gravitational systems in AdS spacetime admits a gauge duality with thermal fields, it is interesting to look how asymptotically spacetimes can be related to thermodynamic systems. With the introduction of charge in the anti-de Sitter spacetime, Kubiznak and Mann \cite{Kubiznak:2012wp} were able to construct an equation of state for the charged AdS black hole analogue to the Van der Waals equation, and to show the existence of several new phase transitions. In this paper we will perform a similar analysis, but using a string cloud background instead of a electromagnetic field, and we will show that in higher dimensions we recover the same Van der Waals equation of state with the same critical parameters.   
 
A cloud of strings (CoS) is the one-dimensional analogous of a cloud of dust and has been introduced, as far as we know, by Letelier \cite{Letelier:1979ej}. For a radial configuration, its static nature can be maintained by the balance between the internal pull of the gravitational field with the external push of a negative pressure due to the string. Despite the fact that the string cloud has been introduced as a configuration for one-dimensional strings, it resembles some others purely energetic field configurations, such as the global monopole in four dimensions \cite{Barriola:1989hx}. The advantage to consider a string cloud configuration is that it can be easily extended from four to arbitrary higher dimensional spacetimes. Thus apart from the physical interpretation of a parameter related to the string cloud, one can apply our whole discussion for the global monopole, for the usual fourth-dimensional spacetime.

In this paper we will consider black holes surrounded by a string cloud both in general relativity and in the framework of $f(R)$ gravity. The latter is today a well-known alternative theory for gravity, that has received a great deal of attention in recent years due to the possibility to explain the current accelerated expansion of the universe by purely geometrical reasons (for reviews, the reader is refereed to \cite{Sotiriou:2008rp,Nojiri:2010wj,Nojiri:2017ncd}). Another reason to study $f(R)$ theories is that it serves as a toy model for an effective theory of gravity, where we can consider the $f(R)$ function as a Fourier expansion in the Ricci scalar. 
\par
The $P-V$ criticality of black holes and its realization as heat engines have already been studied in the framework of $f(R)$ theories \cite{Zhang:2016wek,Chen:2013ce}, but only in the case of a constant Ricci scalar. In this regime, the $f(R)$ gravity is equivalent to general relativity with a cosmological constant, and the theory lacks of interesting features like higher derivatives in the metric field. In the case of the analogous thermodynamic system, only when we consider a non-constant Ricci scalar we are forced to introduce, as we will see, the non-trivial definition of the thermodynamic volume.  

This paper is organized as follows. In section 2, we will briefly introduce the idea of the string cloud and its spacetime metric both in general relativity and in $f(R)$ gravity. In section 3, we will study the black hole as a thermodynamic system in the extended phase space, i.e., with the $P dV$ mechanical work term in general relativity and its realization as a heat engine. In section 4, we will do the same analysis, but in the framework of $f(R)$ gravity. Finally, in section 5 we will present our conclusions.

\section{Black hole with a string cloud in f(R) gravity}

A string cloud is the analogous of a cloud of dust for one-dimensional objects. If one consider a spherically symmetric spacetime, it can be seen as a flow of energy along a radial direction, away from the black hole, starting at its horizon.\footnote{Its presence inside the event horizon is ambiguous, since in principle it is not possible to define a static string configuration inside of it. In any case, we are interested only in the space-time outside the black hole.} In a four dimensional spacetime, it resembles the field of a global monopole, and the effect of the energy-momentum tensor on some background metric is to cause a solid angle deficit. In the case of a higher dimensional spacetime, the effect of the string cloud in the metric is far from trivial, but affects only its radial and time components. The energy-momentum of the string cloud is given by

\begin{equation}
T\indices{^t_t} = T\indices{^r_r} = - \frac{\sigma^2}{r^{d-2}},
\label{energymomentum}
\end{equation}
\noindent
where $\sigma$ is a parameter related to the strength of the cloud, and $d$ is the spacetime dimension. All the other components of the energy-momentum tensor are null (For a discussion on the origin of this energy-momentum tensor, please see \cite{Letelier:1979ej,MoraisGraca:2017nlv}). 

To study the string cloud in $f(R)$ gravity, in principle, we can choose any functional form for the function $f(R)$, and try to solve its field equations. In practice, it is generally not possible to find an exact solution for a non-constant Ricci scalar. The most used method to work in $f(R)$ gravity is to propose a desired solution, in general one that can be analytically written, and then let the equations impose constraints on the $f(R)$ functional form. 

In this paper we will use a $f(R)$ such that $df(R)/dR = 1 + \psi_0 r$, where $\psi_0$ is the parameter that indicates departures from general relativity, and we shall consider $\psi_0r<<1$. One should notice that this restricts the validity of the metric, but as long as we assume $\psi_0$ as a small parameter we can still consider high values of $r$. This constraint imposes that our $f(R)$ theory is a small modification of general relativity. This is not a random guess, but is a constraint imposed by the field equations so that the metric can be found analytically, has a Schwarzschild-like form ($g_{tt}=-g_{rr}^{-1}$) and reduces to the Schwarzschild-Tangherlini-AdS metric in the weak-field limit, i.e., as $\psi_0 \rightarrow 0$ and $\sigma \rightarrow 0$.
\par
For this $f(R)$ function, a spacetime solution can be found for the static spherically symmetric configuration as some of us derived in \cite{MoraisGraca:2017nlv}. In fact, for the metric

\begin{equation}
ds^2 = -A(r) dt^2 + B(r) dr^2 + r^2 \gamma_{ij} dx^i dx^j,
\end{equation}
where $\gamma_{ij}$ is a $(d-2)$-dimensional sphere, and the latin indices $(i,j,...)$ run from $2$ to $d$. Using the $f(R)$ field equations
\begin{equation}
F(R) R_{\mu\nu} - \frac{1}{2}f(R) g_{\mu\nu} - \nabla_{\mu} \nabla_{\nu} F(R) + g_{\mu\nu} \Box F(R) = \kappa^2 T_{\mu\nu},
\label{eom}
\end{equation}
where $F(R)=df/dR$, and the components of the Ricci tensor
\begin{align}
R\indices{^t_t} = - \frac{1}{4} \frac{-A'B'Ar + 2 A''ABr - A'^2Br + 2(d-2) A'BA }{A^2 B^2 r},
\\
R\indices{^r_r} = - \frac{1}{4} \frac{-A'^2 Br + 2A''BAr - A'B'Ar - 2(d-2) B'A^2}{A^2 B^2 r},
\\
R\indices{^i_i} = \frac{1}{2} \frac{-A'Br + B'Ar + 2(d-3)B^2A - 2(d-3)BA}{A B^2 r^2}\, ,
\end{align} 
we derive the metric solution

\begin{equation}
ds^2 = - (1 + A(r)) dt^2 + (1 - A(r)) dr^2 + r^2 \gamma_{i j} dx^i dx^j,
\label{metric1}
\end{equation}
where 
\begin{eqnarray}
\nonumber
A(r) =D_1r^{3-d} + D_2r^2 - \frac{16 \pi G \sigma^2}{d-2}r^{4-d} 
\\
- \frac{2 \psi_0 r}{d-2} + \frac{8 \pi G \sigma^2 \psi_0}{d-3}r^{5-d}.
\label{metric2}
\end{eqnarray}

In this equation $D_1$ and $D_2$ are integration constants related to the black hole mass parameter and a negative cosmological constant, which leads to:

\begin{eqnarray}
\nonumber
A(r) = -\frac{16 \pi GM}{(d-2) \Omega_{d-2}}r^{3-d} + \frac{r^2}{l^2} - \frac{16 \pi G \sigma^2}{d-2}r^{4-d} 
\\
- \frac{2 \psi_0 r}{d-2} + \frac{8 \pi G \sigma^2 \psi_0}{d-3}r^{5-d}
\label{metric2}
\end{eqnarray}

\noindent
where $G$ is Newton's gravitational constant, $\Omega_{d-2}$ is the unitary radius for the $S^{d-2}$-sphere which will be better defined latter, $M$ is a parameter to be related to the black hole mass, and $l^2$ is a squared radius related to the negative cosmological constant $\Lambda$.
\par
Using this solution we can reconstruct the $f(R)$-function. In fact, the metric that we found defines a Ricci scalar $R$, which is a function of the radial coordinate $R=h(r)$. When it is possible to invert this equation as $r=h^{-1}(R)$, we can use our ansatz to recover the function $df(R)/dR=1+\psi_0h^{-1}(R)$. Integrating this equation, we are able to recover the original function $f(R)$ that generated this solution.
\par
In particular, for our case of a string cloud solution, we can analytically derive a function only in four dimensions:
\begin{align}
f(R)=R-2\psi_0\sqrt{9\psi_0^2+16\pi G\sigma^2R+192\pi G\frac{\sigma^2}{l^2}}\nonumber\\
-3\psi_0^2\ln\left(\frac{\sqrt{9\psi_0^2l^2+16\pi G\sigma^2l^2R+192\pi G\sigma^2}-3\psi_0l}{\sqrt{9\psi_0^2l^2+16\pi G\sigma^2l^2R_0+192\pi G\sigma^2}-3\psi_0l}\right)\nonumber\\
+3\psi_0^2\ln\left(\frac{\sqrt{9\psi_0^2l^2+16\pi G\sigma^2l^2R+192\pi G\sigma^2}+3\psi_0l}{\sqrt{9\psi_0^2l^2+16\pi G\sigma^2l^2R_0+192\pi G\sigma^2}+3\psi_0l}\right)\nonumber\\
+3\psi_0^2\ln\left(\frac{l^2R+12}{l^2R_0+12}\right),
\end{align}
where $R_0$ is an integration constant. For dimensions greater than four approximative methods should be employed for deriving an effective $f(R)$ function.
\par
For the case of the black hole with a string cloud in general relativity, we assume that when $\sigma \rightarrow 0$ we obtain the Schwarzschild-Tangherlini-AdS metric, not only in the weak field limit. This means that, for general relativity, the spacetime for the black hole with a string cloud will be given by the metric

\begin{equation}
ds^2 = - B(r) dt^2 + B(r)^{-1} dr^2 + r^2 \gamma_{i j} dx^i dx^j, 
\end{equation}
where

\begin{equation}
B(r) = 1 - \frac{16 \pi GM}{(d-2) \Omega_{d-2}}r^{3-d} + \frac{r^2}{l^2} - \frac{16 \pi G \sigma^2}{d-2}r^{4-d}. 
\end{equation}

For a detailed discussion on how to obtain the above metrics, and for a study on the classical thermodynamics of the black hole string cloud in $f(R)$ gravity, the reader is referred to \cite{MoraisGraca:2017nlv}. 

In the following section we will study the $P-V$ criticality and the realization of the black hole as a heat engine in general relativity. 

\section{Black hole thermodynamics in general relativity}\label{grcase}

In this section we will briefly review the thermodynamics of black holes in general relativity and apply it for the case of a black hole with a string cloud. In the next section, we will extend our study for $f(R)$ gravity. 
\par
The radius of the event horizon, also called the horizon radius, $r_+$, satisfies the following equation

\begin{equation}
1 - \frac{16 \pi GM}{(d-2) \Omega_{d-2}}r_+^{3-d} + \frac{r_+^2}{l^2} - \frac{16 \pi G \sigma^2}{d-2}r_+^{4-d} = 0, 
\end{equation}
which allows us to write the integration constant, $M$, as 

\begin{equation}
M = \frac{(d-2) \Omega_{d-2} r_+^{d-3}}{16 \pi G} \left(1 + \frac{r_+^2}{l^2} - \frac{16 \pi G \sigma^2 r_+^{4-d}}{d-2}\right).
\label{integrationConstant}
\end{equation} 

By our convention, the ADM mass of the system is related to the integration constant by $M_{adm} = 4\pi M / \Omega_{d-2}$, where

\begin{equation}
\Omega_{d-2}=\frac{2\, \pi^{(d-1)/2}}{\Gamma((d-1)/2)}.
\end{equation}

As can be seen, in four dimensions the ADM mass is equal to the integration constant M, and for an arbitrary dimensional we have

\begin{equation}\label{mass-parameter}
M_{adm} = \frac{1}{4Gl^2} \frac{1}{r_+^{3-d}} [(d-2)(r_+^2+l^2) - 16 \pi G \sigma^2 l^2 r_+^{4-d}].
\end{equation}

The Hawking temperate can be found using the \textit{Euclidean trick}, and it can be identified as

\begin{widetext}
\begin{equation}
T =  \frac{\hbar}{\kappa_B \Omega_{d-2}}\left[\frac{dg(r)}{dr}\right]_{r=r_+} = \frac{\hbar}{\kappa_B \Omega_{d-2}} \frac{1}{l^2 r_+} [(d-3)l^2 + (d-1) r_+^2 - \frac{16 \pi G \sigma^2 l^2}{d-2}r_+^{4-d}],
\label{temperature}
\end{equation}
\end{widetext}

\noindent
where $\hbar$ is the reduced Planck constant and $\kappa_B$ is the Boltzmann constant. The Bekenstein-Hawking entropy is related to the area of the black hole, and is given by

\begin{equation}
S = \frac{\kappa_B}{4G\hbar} \Omega_{d-2} r_+^{d-2},
\label{entropy}
\end{equation}

\noindent
and we are now ready to express the first law of thermodynamics. 
\par
As was shown in the last section, the black hole mass $M$ and the cosmological constant $l$ were recognized from integration constants when we derived the metric from the field equations. Therefore, the cosmological constant $l$ contributes to the metric as a parameter of the black hole geometry, just as a the mass $M$. Then it is relevant to consider not just $M$, but also $l$ as varied when analyzing the black holes mechanics/thermodynamics. A similar role of the cosmological constant was also contemplated in the seminal paper \cite{Teitelboim:1985dp} on this subject (see also \cite{Bunster:2014cna} for the treatment of the cosmological constant as a dynamical field) Allowing $l$ to vary, the differential of the ADM mass can be written as

\begin{equation}
dM_{adm} = T dS - \frac{1}{4 G r_+^{3-d}}2 (d-2) \frac{r_+^2}{l^3} dl,
\label{dM}
\end{equation}
and thus can be compared with the thermodynamic definition of enthalpy, $dH = TdS + VdP$, where $V$ is the thermodynamic volume of the system and $P$ is its pressure. Let us note that, in the absence of the cosmological constant, one simply obtains $dM_{adm} = T dS$. Equation (\ref{dM}) allows us to identify the ADM mass with the enthalpy, and, defining the $(d-1)$-dimensional volume as

\begin{equation}
V_{d-1} = \frac{\pi^{(d-1)/2}}{\Gamma((d+1)/2)} r_+^{d-1},
\label{volume}
\end{equation} 

\noindent
one can identify the pressure of the thermodynamic system as

\begin{equation}
P = \frac{d-2}{4G} \frac{\Gamma((d+1)/2)}{\pi^{(d-1)/2}} \frac{1}{l^2}.
\label{pressure}
\end{equation}

Another motivation for this identification follows from recognizing that that a negative cosmological constant induces a positive vacuum pressure in the spacetime, as pointed out in the review \cite{Kubiznak:2016qmn}.
\par
For $d=4$, we get the usual identification $P = 3/8\pi G l^2$. Using the relations (\ref{volume}) and (\ref{pressure}), we can rewrite (\ref{temperature}) as an equation of state, given by

\begin{widetext}
\begin{equation}
P = \frac{(d-2)}{(d-1)} \frac{\Gamma(\frac{d+1}{2})^{(d-2)/(d-1)}}{4G \pi^{(d-2)/2}} \left[ \frac{\kappa_B \Omega_{d-2} T}{\hbar V^{1/(d-1)}}- \frac{(d-3) \sqrt{\pi}}{\Gamma(\frac{d+1}{2})^{1/(d-1)} V^{2/(d-1)}} + \frac{16 \pi G \sigma^2}{(d-2)} \frac{\pi^{(d-3)/2}}{\Gamma(\frac{d+1}{2})^{(d-3)/(d-1)} V^{(d-2)/(d-1)}} \right]\, .
\label{equationOfState}
\end{equation}

We can also write the temperature as a function of the entropy and the pressure, as

\begin{equation}\label{tempgr}
T = \frac{\hbar}{\kappa_B \Omega_{d-2}}\frac{1}{\tilde{S}^{1/(d-2)}}\left[d-3+(d-1)\tilde{S}^{2/(d-2)}\tilde{P} - \frac{16 \pi G\sigma^2}{d-2} \tilde{S}^{(4-d)/(d-2)}\right],
\end{equation}

\noindent
where $\tilde{S} \doteq 4 G\hbar S / \kappa_B \Omega_{d-2}$ and $\tilde{P} \doteq 4 G \pi^{(d-1)/2} P / (d-2) \Gamma((d+1)/2)$. With this equation, we can compute the heat capacity as $C = T \partial S / \partial T$, obtaining the following result 

\begin{equation}
C = \frac{\kappa_B \Omega_{d-2}}{4 G \hbar} (d-2)\tilde{S} \left(1 - \frac{\hbar (d-1)}{\kappa_B \Omega_{d-2}} \tilde{S}^{1/(d-2)}\frac{\partial \tilde{P}}{\partial T}\right) \frac{[(d-1)\tilde{P}\tilde{S}^{d/(d-2)} + (d-3)\tilde{S} - 16 \pi \sigma^2 \tilde{S}^{2/(d-2)}/(d-2)]}{[(d-1)\tilde{P}\tilde{S}^{d/(d-2)} - (d-3)\tilde{S} + 16 (d-3) \pi \sigma^2 \tilde{S}^{2/(d-2)}/(d-2)]}.
\end{equation}
\end{widetext}

If one calculates $\partial \tilde{P} / \partial T$ at constant volume, one will find that the above quantity vanishes. This means that the heat capacity at constant volume, $C_V$, is null. This result is a consequence of the fact that both the volume and entropy are functions of the horizon radius only, as will be clear when we study the same object in $f(R)$ gravity. Thus for any process involving the system, isochorics are also adiabatics, a feature that will be explored when we define the cyclic process of our heat engine. 

To follow up our analysis, we will work with the functional form of the pressure depending on the horizon radius and the temperature. This choice is motivated by the fact that the quantity $\nu=2\ell_P^2r_+$ plays the role of the specific volume of the system, i.e, it is the thermodynamic volume divided by the number of degrees of freedom of the horizon $N=A/\ell^2_P$, where $\ell_P$ is the Planck length. In the major part of the literature that compares the black hole behavior with a Van der Waals fluid, this quantity is used to represent the corresponding specific volume. Using the definition of $\Omega_{d-2}$ and the recurrence relation $\Gamma(\alpha+1)=\alpha\Gamma(\alpha)$, one can rewrite the pressure (\ref{equationOfState}) as

\begin{widetext}

\begin{equation}
P(r_+,T)=\frac{(d-2)}{4G(d-1)}\left[\frac{(d-1)k_B}{\hbar}\frac{T}{r_+}-\frac{(d-3)}{r_+^2}\pi^{\frac{1-d}{2}}\Gamma\left(\frac{d+1}{2}\right)+\frac{16\pi G \sigma^2}{d-2} \pi^{\frac{1-d}{2}}\Gamma\left(\frac{d+1}{2}\right)r_+^{2-d}\right]\, .
\label{equationOfState2}
\end{equation}

\end{widetext}

The critical temperature, $T_c$, and radius, $r_c$, can be found by solving the following system of equations
\begin{subequations}\label{critical_conditions}
\begin{align}
\left(\frac{\partial P}{\partial r_+}\right)\Big |_T=0,\\
\left(\frac{\partial^2 P}{\partial r_+^2}\right)\Big |_T=0,
\end{align}
\end{subequations}

\noindent
where it is straightforward to find that
\begin{align}
T_c=\frac{\hbar\,}{k_B}\frac{(2d-8)}{d-1}(8G\sigma^2)^{\frac{1}{4-d}}\pi^{\frac{(d-2)(d-3)}{2(4-d)}}\Gamma\left(\frac{d+1}{2}\right)\, ,\label{tc1}\\
r_c=\left(8\pi G \sigma^2\right)^{\frac{1}{d-4}}\, \label{rc1}.
\end{align}
As can be inferred from the equations above, there is no criticality for $d=4$, i.e., the critical temperature and radius are not defined, which is a behavior that is in contradiction with previous cases like the Reissner-Nordstrom-AdS (RN-AdS) black hole \cite{Kubiznak:2012wp}. We should also note that for $d=3$, the critical temperature is negative. The most interesting cases occur for $d>4$, where the behavior of the thermodynamic quantities are similar to previous cases and the criticality can be fully realized.
\par
We also draw attention to the interesting relation between the critical quantities, given by
\begin{equation}
\frac{P_c\, r_c}{T_c}=\frac{(d-3)}{8}\frac{k_B}{G\hbar}\, ,
\end{equation}
\noindent
where $P_c\doteq P(r_c,T_c)$. This is a universal relation, which is valid for any string cloud, since it does not depend on $\sigma$. For $d=6$, one find
\begin{equation}
\frac{G\hbar}{k_B}\frac{P_c\, r_c}{T_c}{\Big |}_{d=6}=\frac{3}{8},
\end{equation}

\noindent
which turns out to be the same numerical value found when analyzing a Van der Waals fluid by a RN-AdS black hole system. This happens because only in $d=6$, the string cloud can emulate the contribution due to the electronic charge in four-dimensions, which is a $r_+^{-4}$ contribution in the equation of state for the pressure (\ref{equationOfState2}) (to be compared with Eq.(3.18) in \cite{Kubiznak:2012wp}). 
\par
This duality between four and six dimensions is useful for the study of criticality, when considering the ADM mass as the enthalpy of the system, but this would be also the case if we work with the Gibbs free energy $G(T,P)=H-ST$ and the free energy $F(T,V)=G-PV$ for finding the critical exponents. However, this is not a perfect duality, since the ADM masses are not equivalent in these distinct cases, which will lead to a distinct behavior of the efficiency of this black hole as a heat engine for the square circle $P-V$ diagram, as we will see in the next section, in Eq.(\ref{efficiencySquaredHE}).
 \par
\begin{figure}[ptb]
\includegraphics[scale=0.42]{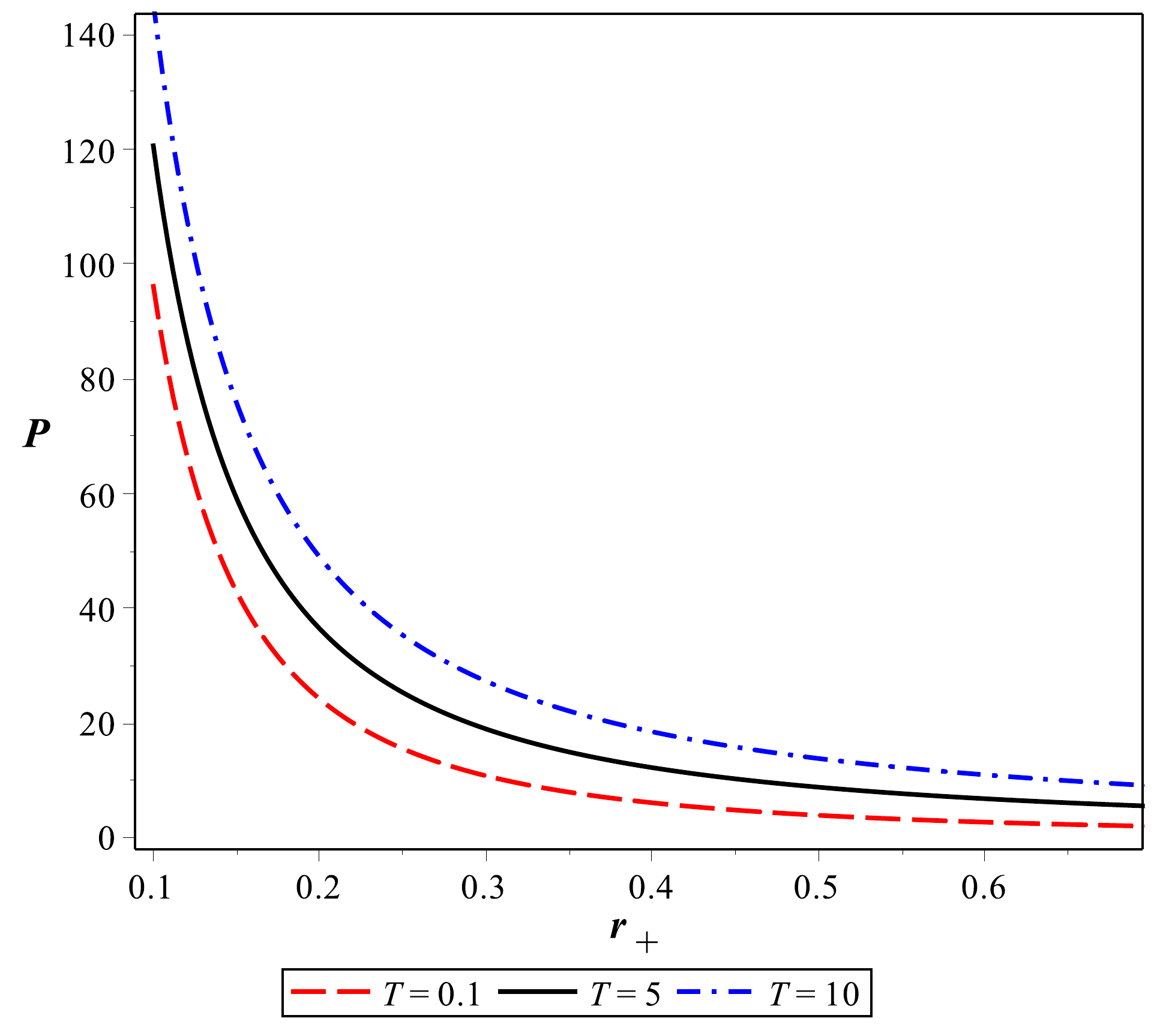}
\caption{$P-r_+$ diagram in the four-dimensional case, for $\sigma=1$.}
\label{4dPV}
\end{figure}
\begin{figure}[ptb]
\includegraphics[scale=0.42]{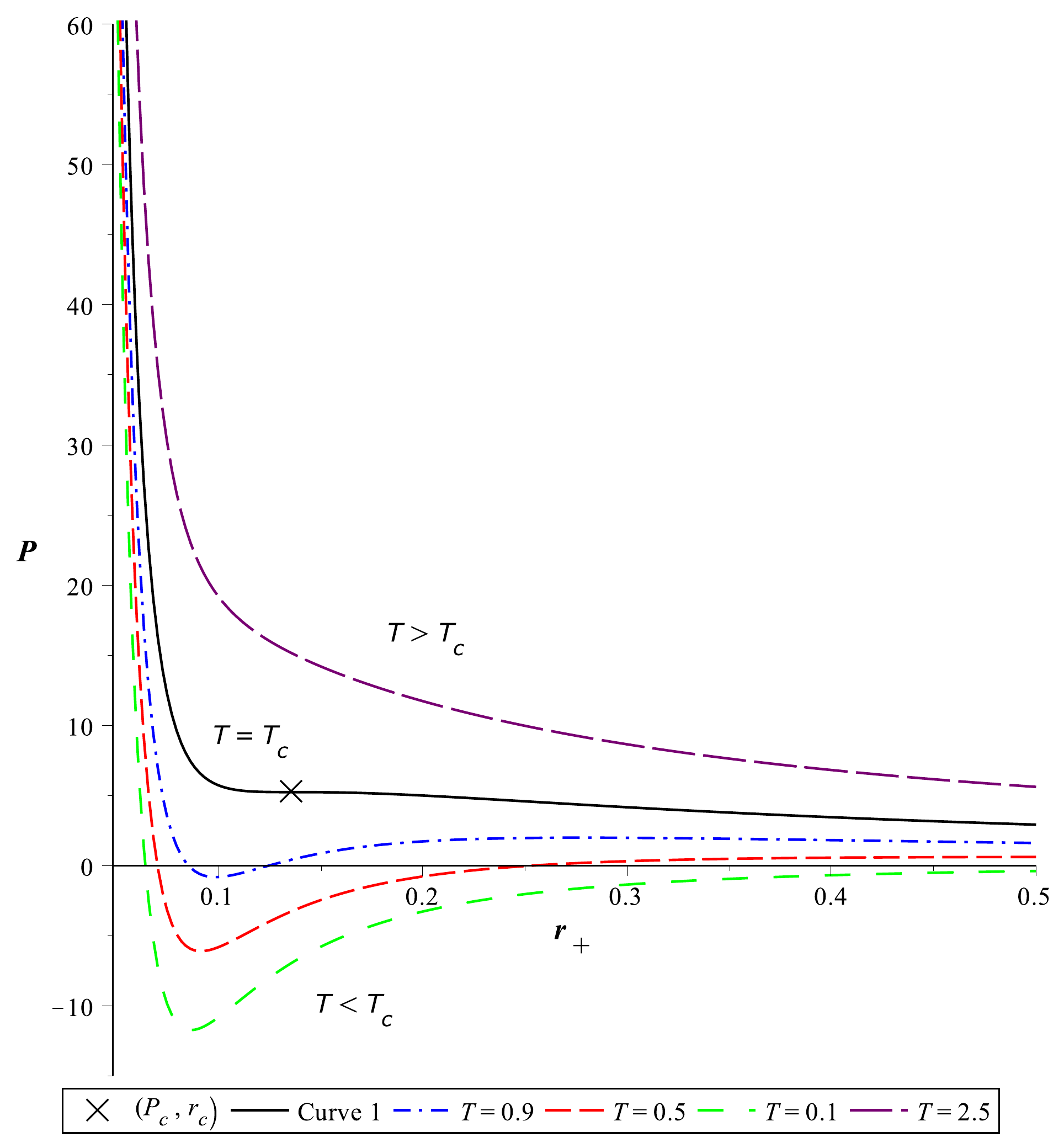}
\caption{$P-r_+$ diagram in the seven-dimensional case, for $\sigma=10^{-2}$.}
\label{7dPV}
\end{figure}

We illustrate the $P-r_+$ diagrams in Figs.(\ref{4dPV}), (\ref{7dPV}) (we set $\hbar=k_B=G=1$). As can be seen, there are no critical values for d=4, which implies that the diagram always follow the same behavior, i.e., $\partial P/\partial r_+$ at constant temperature is always negative. For a van de Waals fluid, $T_c$ means the temperature above which it is not possible to distinguish liquid and gas phases; for the black hole chemistry the analogy consists in translating liquid/gas to small/large black hole. Therefore, in our case of a string cloud configuration, in $d=4$, due to the absence of a critical temperature, the thermodynamic behavior is the same for small and large black holes.
\par
For $d>4$, there exist critical quantities, which implies that below $T_c$, there are regions in the $P-r_+$ diagram that allow for distinctions between the different ``phases'' of small/large black holes. For instance, in Fig.(\ref{7dPV}), we depicted the diagram for seven dimensions, in which can be seen the qualitative behavior of the pressure $P(r_+)$ for $T<T_c$, where a phase transition occurs. We again remark that the qualitative behavior is similar to the RN-AdS black hole analyzed in \cite{Kubiznak:2012wp}. For a discussion of the phase transitions based on the non-extended phase space, i.e., without the $P dV$ working term, the reader is refereed to \cite{Dey:2017xty}

\subsection{Heat engines}

The subject of thermodynamics has been developed as a tool to study the behavior and efficiency of heat engines, i.e., machines that produce work using some substance as its \textit{fuel}. In this framework it is natural that, after considering black holes as thermodynamic systems, one can consider such system as a heat engine, where the substance used by such engine is defined by its equation of state, given by Eq.(\ref{equationOfState}) in the case of the present paper. 

\begin{figure}[]
\centering
\begin{tabular}{@{}cc@{}}
\includegraphics[scale=0.60]{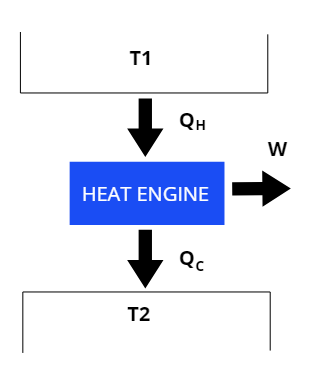}
\end{tabular}
\caption{A 2T heat engine}
\label{fig:heatEngine}
\end{figure}

The most natural heat engines are the ones working between two reservoirs of temperature, as sketched in figure (\ref{fig:heatEngine}), where $Q_H$ stands for the amount of heat extracted by the warmer reservoir $T_1$, and $Q_C$ is the amount of heat rejected on the colder reservoir $T_2$. The main function of the engine is to produce some positive work, represented by the letter $W$.

It has been proved by Carnot that any other heat engine has lower or at least equal efficiency compared with a heat engine that operates a reversible cycle, such as the one named Carnot cycle, constructed by two isothermals and two adiabatics.  As is well-know, the efficiency of a heat engine is defined by

\begin{equation}
\eta \doteq \frac{W}{Q_H},
\label{efficiency}
\end{equation} 

\begin{figure}[]
\centering
\begin{tabular}{@{}cc@{}}
\includegraphics[scale=0.60]{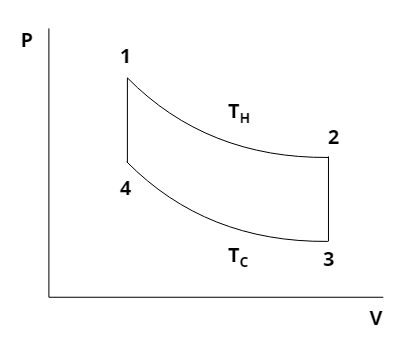}
\end{tabular}
\caption{A Carnot cycle for null $C_V$}
\label{fig:carnotCycle}
\end{figure}

\noindent
where $W$ and $Q_H$ has already been introduced. For the case of black hole heat engines, since isochorics are adiabatics, the Carnot cycle for our static black hole is simpler and can be  represented by figure (\ref{fig:carnotCycle}). One feature of the Carnot engine is that its efficiency depends only on the temperature of the reservoirs, and it is given by

\begin{equation}
\eta = 1 - \frac{T_C}{T_H},
\label{efficiencyCarnotHE}
\end{equation}

\noindent
where $T_C$ and $T_H$ are the temperatures of the two reservoirs, considering that $T_H > T_C$. Calculating the efficiency of the black hole as a Carnot heat engine is straightforward, since we can use equation (\ref{efficiencyCarnotHE}) with the temperature of the black holes. 

\begin{figure}[]
\centering
\begin{tabular}{@{}cc@{}}
\includegraphics[scale=0.60]{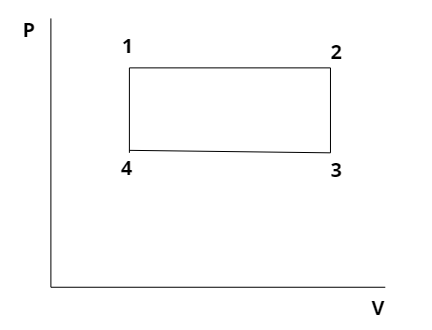}
\end{tabular}
\caption{A square cycle in the p-V diagram}
\label{fig:squareCicle}
\end{figure}

For an arbitrary cycle, the efficiency should be calculated by Eq.(\ref{efficiency}), and hardly it will be possible to find an analytic formula, given the complexity of the equation of state (\ref{equationOfState}). In his pioneering paper \cite{Johnson:2014yja}, Johnson has proposed a simple enough cycle so one could calculate an exact expression for the efficiency. This cycle is composed by two adiabatics/isochorics and two isobarics, as represented in figure (\ref{fig:squareCicle}). The work done by this heat engine is the area of the rectangle $1 \rightarrow 2 \rightarrow 3 \rightarrow4$, and the heat absorbed by the system along the isobar $1 \rightarrow 2$ can be easily found using the first law of thermodynamics,

\begin{equation}
dH = \delta Q + V dP.
\end{equation}

\noindent
Since the pressure is constant, we have that the heat absorbed is simply the enthalpy, that is the ADM mass of the black hole. For this reason, the efficiency of the rectangular cycle is given by

\begin{equation}
\eta = \frac{W}{Q_A} = 1 - \frac{Q_R}{Q_A} = 1 - \frac{M_3 - M_4}{M_2 - M1},
\label{efficiencySquaredHE}
\end{equation}

\noindent
where $Q_R$ is the heat emitted by the system, and $Q_A$ the heat absorbed by the system. We should stress out the fact that this is not a heat engine operating between two reservoirs. As we mentioned before, due to the dimension-dependence of the ADM mass from Eq.(\ref{mass-parameter}), we see that the duality between the four-dimensional RN-AdS and six-dimensional CoS-AdS black holes is not perfect, since the efficiency of this square cycle allows us to distinguish between these configurations.
\par
Given these considerations, we are now able to study the heat engine in more details. We will not explicitly write the formulas for equations (\ref{efficiencyCarnotHE}) and (\ref{efficiencySquaredHE}) since it can be easily found if one substitute the values for mass and temperature given by Eqs. (\ref{mass-parameter})  and (\ref{temperature}), respectively. Instead, we will plot the Carnot efficiency and the squared cycle efficiency as functions of the string cloud parameter $\sigma$. Our aim is to analyze the effect of the introduction of a string cloud in the efficiency of the black hole as a heat engine.

\begin{figure}[]
\centering
\begin{tabular}{@{}cc@{}}
\includegraphics[scale=0.40]{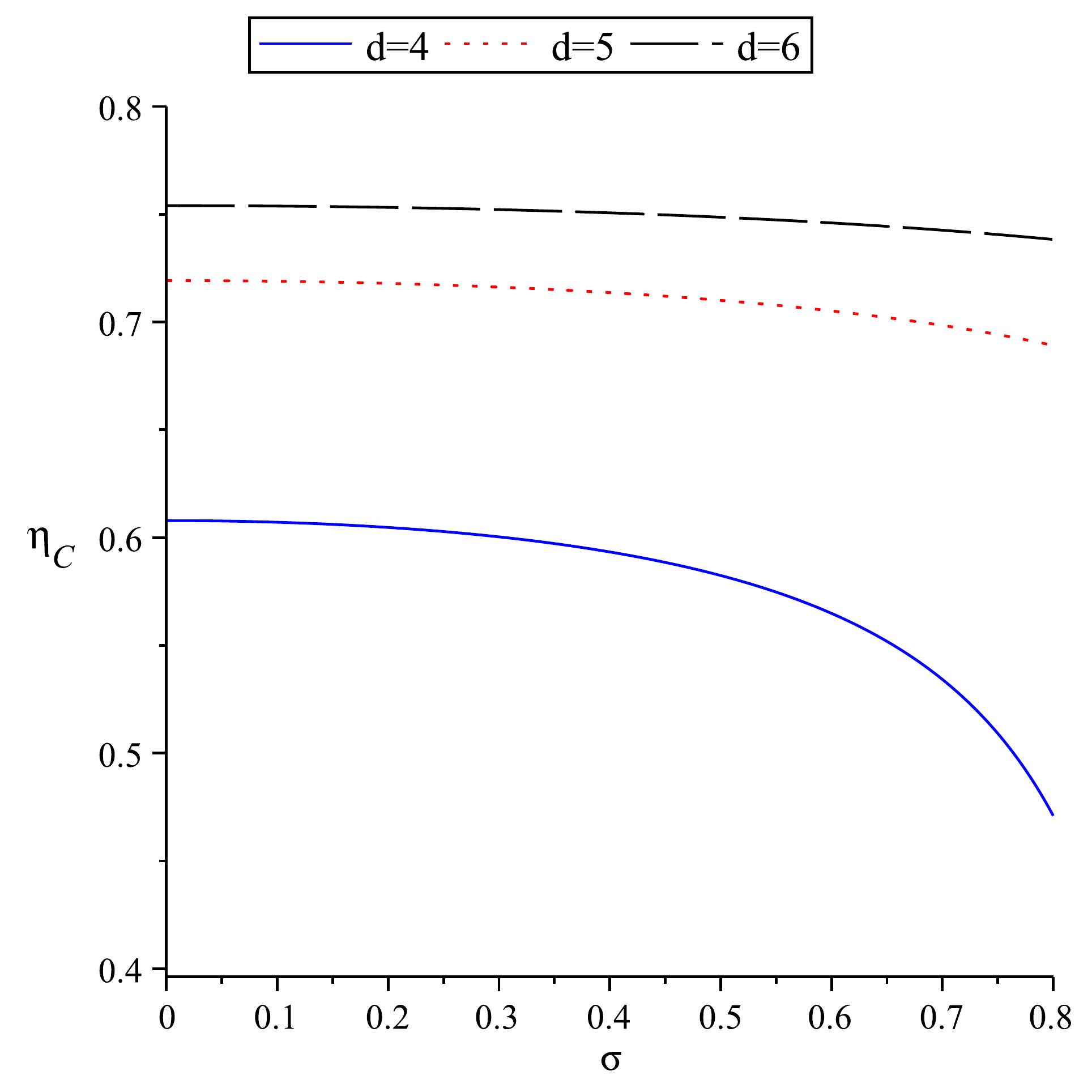}
\end{tabular}
\caption{Efficiency of the Carnot cycle as a function of the string cloud parameter $\sigma$}
\label{fig:effCarnotGR}
\end{figure}

In figure (\ref{fig:effCarnotGR}), we analyze the efficiency of the Carnot cycle for $\tilde{S}_H = 1, \tilde{P}_H = 8, \tilde{S}_C = 10$ and $\tilde{P}_C = 1$, where the subscripts $H$ and $C$ stands for \textit{Hot} and \textit{Cold}. All other constants are set to be equal unity. The choice criteria was that both the temperatures should be positive, and $T_H$ greater than $T_C$. As can be seen in the mentioned figure, the efficiency of the heat engine decreases as the string parameter increases, for dimensions four to six. It is not possible, however, to conjecture that this behavior will be valid for all temperatures, given the relation between temperature, dimension and string parameter. 

\begin{figure}[]
\centering
\begin{tabular}{@{}cc@{}}
\includegraphics[scale=0.40]{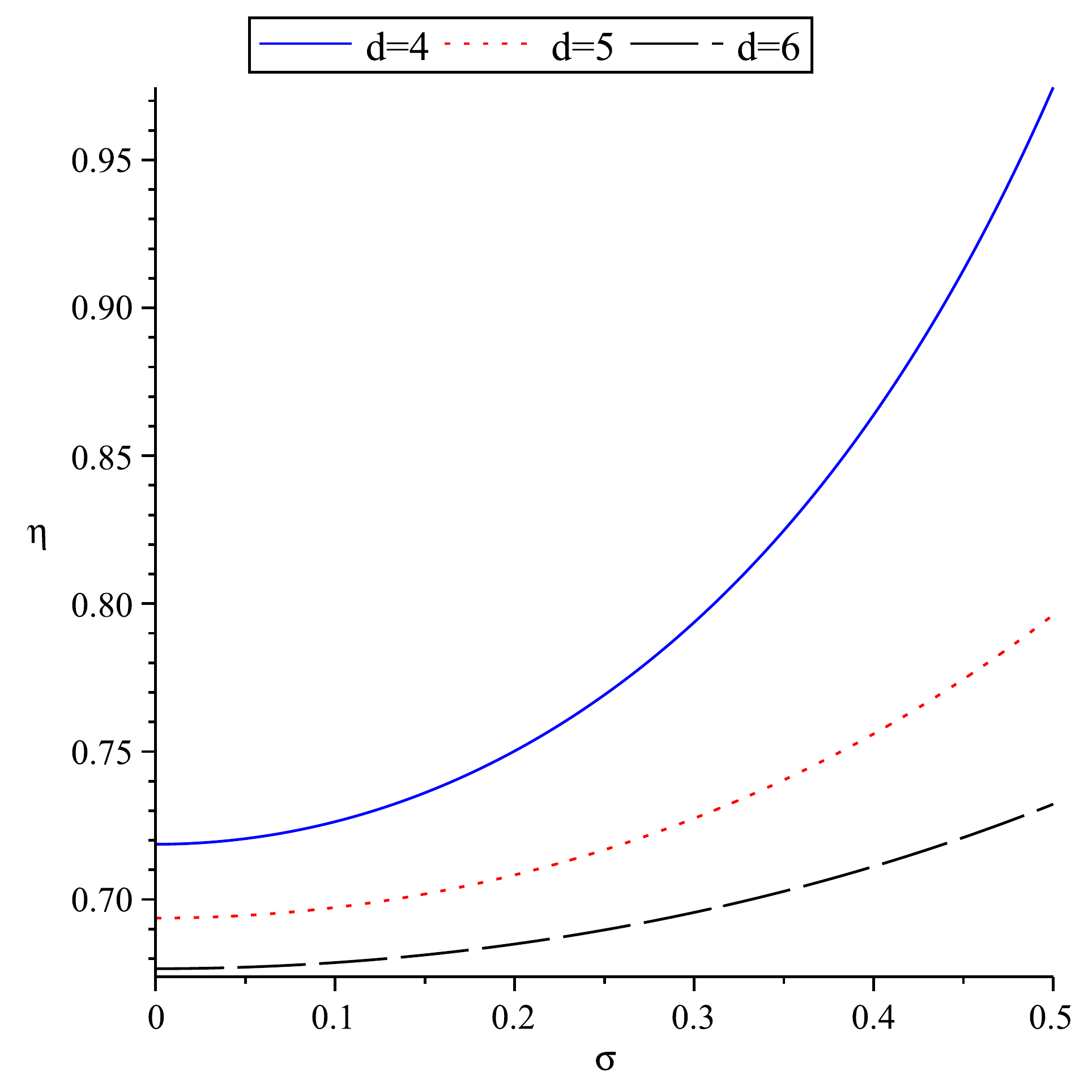}
\end{tabular}
\caption{Efficiency of the squared cycle as a function of the string cloud parameter $\sigma$}
\label{fig:effSquaredGR}
\end{figure}

In figure (\ref{fig:effSquaredGR}), we analyze the efficiency of the cycle depicted in figure (\ref{fig:squareCicle}), for the following values: $\tilde{S}_1 = \tilde{S}_4 = 1, \tilde{P}_1 = \tilde{P}_2 = 4, \tilde{S}_2 = \tilde{S}_3 = 3 $ and $\tilde{P}_3 = \tilde{P}_4 = 1$. All the other parameters are set to be equal unity, and the choice criteria was that $M_3 > M_4$ and $M_2 > M_1$. For this cycle, we found that the efficiency of the engine increases as long as we increase the parameter $\sigma$. It actually appears that it will be greater than unity for $\sigma$ larger than $0.25$, but this value is the limit for all masses to be positive. Once again, it is not possible to conjecture that the efficiency of the heat engine will be increased for all masses and dimensions, due to a increase of the string cloud parameter $\sigma$, but we can affirm so at least for a range of parameters.

\section{Black hole thermodynamics in f(R) gravity}

A more detailed account of the thermodynamics of a black hole with a string cloud, in $f(R)$ gravity, has already been studied in \cite{MoraisGraca:2017nlv}. However, this previous study has been performed in the non-extended phase space, i.e., without a mechanical work term, and by this reason neither a proper definition of pressure nor the $P-V$ criticality have been studied. We will perform this study in this section.
\par
The extended phase space approach for black hole thermodynamics for a particular example of $f(R)$ gravity has been recently studied in \cite{Zhang:2016wek,Chen:2013ce}, however these papers only considered the case of a constant Ricci scalar, which relates to the cosmological constant. In our case, we study the effects of modified gravity for a coordinate-dependent contribution coming from the $f(R)$-function. As we will see, non-trivial contributions will emerge from our approach, like the need for a redefinition of the thermodynamic volume (an issue that also occurs in other theories of gravity, as can be seen in \cite{Kubiznak:2016qmn}), and a the need to combine the string cloud and $f(R)$ parameters, $\sigma$ and $\psi_0$ for achieving maximum efficiency of the heat engine.
\par
It has been argued (see Ref.\cite{Faraoni:2010yi} and references therein) that, for $f(R)$ gravity, the entropy of black holes is modified by a factor proportional to the derivative of the $f(R)$ functional form in respect to the Ricci scalar. Thus, we should modify our previous definition of entropy, Eq.(\ref{entropy}), to the following definition

\begin{equation}
S = \frac{\kappa_B}{4 G \hbar} \Omega_{d-2} r_+^{r-2} f'(R),
\end{equation}
\noindent
where $f'(R) = df(R)/dR$. 

The temperature of the black hole has the same definition as in general relativity. Thus, for the metric given by equations (\ref{metric1}) and (\ref{metric2}), in the leading order, one can find the temperature as

\begin{eqnarray}
\nonumber
T =  \frac{\hbar}{\kappa_B \Omega_{d-2}} \frac{1}{l^2 r_+} [(d-3)l^2 + (d-1) r_+^2 
\\
- \frac{16 \pi G \sigma^2 l^2}{d-2}r_+^{4-d} - 2 \psi_0 l^2 r_+].
\label{temperaturefR}
\end{eqnarray}

To proceed, we now argue that the first law is valid, providing us with a method to find the enthalpy of the system as

\begin{equation}
dH = T dS|_{p},
\end{equation}
\noindent
were the pressure is kept constant. Integrating the above equation one finds that

\begin{eqnarray}
\nonumber
H = \frac{1}{4Gl^2} [(d-2)l^2 r_+^{d-3} - 16 \pi G l^2 \sigma^2 r_+ 
\\
\nonumber
+\frac{(d^2 - 6d + 7)}{d-2} l^2 \psi_0 r_+^{d-2} + (d-2) r_+^{d-1} 
\\
+ \frac{(d-1)^2}{d} \psi_0 r_+^d].
\label{enthalpyFR}
\end{eqnarray}

\noindent
The pressure of the system can be defined keeping the volume constant and taking the differential of the above equation, in an analogous way as we did for general relativity. However, unlike general relativity, the definition of the thermodynamic volume will not be so straightforward. The differential of (\ref{enthalpyFR}) for constant horizon radius is given by

\begin{equation}
dH|_{r^+} = - \frac{2}{4Gl^3}\left( (d-2) r^{d-1} + \frac{(d-1)^2}{d} \psi_0 r^d\right) dl,
\label{enthlapyFR2}
\end{equation}
\noindent
and it can be seen that or the pressure or the thermodynamic volume (or both) should incorporate the parameter $\psi_0$. 

In previous works on $P-V$ criticality in black holes in $f(R)$, it has been argued that the pressure should be redefined in the same way as the entropy, i. e., $P \propto f'(R)$ (see \cite{Zhang:2016wek,Chen:2013ce}). However, this was possible only because such works deal with the case where the Ricci scalar is a constant. In such case, the volume can be defined as the usual volume $V_{d-1} \propto r^{d-1}$, and the pressure can be a function only of the parameter $l$. This happens because $f(R)$ gravity with a constant Ricci scalar is equivalent to general relativity with a cosmological constant. In a general $f(R)$ theory of gravity, it is not possible to do both choices at the same time, i.e., one must give up to define the thermodynamic volume as the naive volume proportional $r_+^d$ or to define the pressure as a pure function of the length parameter $l$. This is not an issue only for $f(R)$ gravity. Even for a Kerr-AdS black hole in general relativity, its thermodynamic volume should be redefined to include its angular momentum (for a deeper discussion, see \cite{Kubiznak:2016qmn}). 

For consistency, the pressure of the system should depend only on the cosmological parameter $l$, and not on the horizon radius, since the entropy (that depends only on $r_+$) and the pressure are supposed to be independent variables. Thus, the thermodynamic volume should be redefined to include the parameter $\psi_0$. To do so, we maintain the definition of the pressure as

\begin{equation}
P = \frac{(d-2)}{4G} \frac{\Gamma((d+1)/2)}{\pi^{(d-1)/2}} \frac{1}{l^2},
\end{equation} 

\noindent
so that, to be able to write equation (\ref{enthlapyFR2}) as the mechanical displacement $ V_{d-1} dP$, one must define the thermodynamic volume as

\begin{equation}
V_{d-1} = \frac{\pi^{(d-1)/2}}{\Gamma((d+1)/2)} r_+^{d-1} \left( 1 + \frac{(d-1)^2}{d(d-2)} \psi_0 r_+ \right).
\end{equation}

Despite the fact that we are working in one particular $f(R)$ theory of gravity, it is expected that the thermodynamic volume of a generic $f(R)$ gravity will be modified in relation to is \textit{naive} definition of volume. Due to this reason, it is no longer a straightforward procedure to find the equations $P(V,T)$ and $T(S,P)$. It is possible, however, to find an equation of state of the form $P(r_+,T)$, and to use it to study the criticality of the system.

\par
\begin{figure}[ptb]
\includegraphics[scale=0.44]{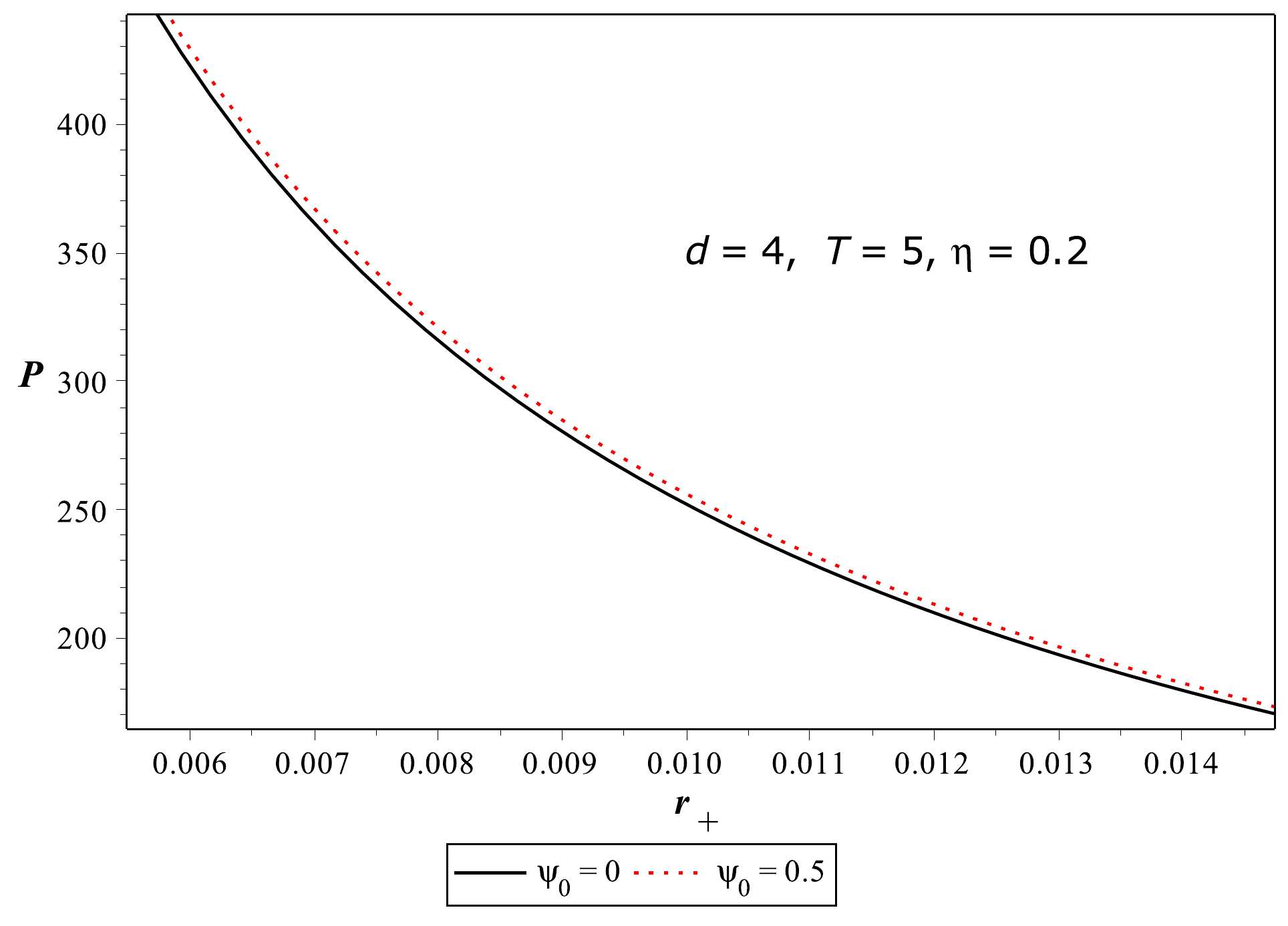}
\caption{$P-r_+$ diagram in the four-dimensional case for $\sigma=0.2$.}
\label{4df(r)}
\end{figure}

With these informations, we can calculate the equation of state for the pressure $P(r_+,T)$ with the $f(R)$ correction as
\begin{equation}\label{EqOfStatef(r)}
P^{f(R)}(r_+,T)=P^{\text{GR}}(r_+,T)+\frac{(d-2)}{4G}\frac{\psi_0}{r_+}\pi^{\frac{1-d}{2}}\Gamma\left(\frac{d-1}{2}\right),
\end{equation}
\noindent
where $P^{\text{GR}}$ is the pressure calculated from general relativity (\ref{equationOfState2}). From the equation above, one can see that for $\psi_0>0$, the pressure increases when compared to general relativity, specially for small black holes. This contribution is proportional to $r_+^{-1}$, which adds up to the temperature term, as can be inferred from Eq.(\ref{equationOfState2}). This implies that also in the $f(R)$ case, critical quantities do not exist in four dimensions, since in order to exist solutions to the criticality conditions given by Eq.(\ref{critical_conditions}), it is mandatory for the pressure to exhibit terms other than $r_+^{-1}$ and $r_+^{-2}$.
 We draw the behavior of the $P-r_+$ diagram in four dimensions in Fig.(\ref{4df(r)}); where it can be seen that the $f(R)$ correction is highly suppressed by the general relativity contribution. This behavior is found also for arbitrary dimensions.

\par
As in the previous case, the criticality conditions (\ref{critical_conditions}) can be solved for $d>4$, as
\begin{eqnarray}
T_c^{f(R)}=T_c^{\text{GR}}-\pi^{\frac{1-d}{2}}\frac{\hbar}{k_B}\psi_0\, \Gamma\left(\frac{d-1}{2}\right),\\
r_c^{f(R)}=r_c^{\text{GR}},
\end{eqnarray}
\noindent
where $T_c^{\text{GR}}$ and $r_c^{\text{GR}}$ are given by Eqs.(\ref{tc1}) and (\ref{rc1}), respectively. The critical radius do not depend on $\psi_0$ and the critical temperature is reduced, although the isotherm $P(r,T_c)$ of the critical temperature does not depend on $\psi_0$ and is the same as the one of general relativity, i.e., the thermal boundary between the distinguishability and indistinguishability of black hole phases (small/large) is unaffected by $f(R)$ gravity. 
\par
The $\psi_0$ contribution is suppressed by the other terms in Eq.(\ref{EqOfStatef(r)}), hence there is no significant difference between the diagrams for $f(R)$ and general relativity (for instance depicted in Fig.(\ref{7dPV}) for $d=7$). Then we ask the reader to refer to the graphs of section \ref{grcase} for a qualitative outlook of the pressure behavior in the $f(R)$ case too. As we will see, this will not be the case for the analysis of the heat engine.
\par
We also calculate the corrected relation between the critical quantities $(P_c,T_c,R_c)$, as
\begin{equation}
\left(\frac{P_cr_c}{T_c}\right)^{f(R)}=\left(\frac{P_cr_c}{T_c}\right)^{\text{GR}}+\frac{(d-3)k_B[(8G)^{5-d}\pi\sigma^{2}]^{\frac{1}{d-4}}}{\hbar(d-4)} \psi_0.
\end{equation}
\noindent
In this case, there is no a universal relation for any string cloud configuration, since there is an explicit dependence on the parameter $\sigma$.

\subsection{Heat engine}

One of the special features of black holes as heat engines is that isochorics are equal to adiabatics. We have shown this important result for general relativity, and now we discuss it in the framework of $f(R)$ gravity. As we do not have an explicit equation of the form $T(S,P)$, we can calculate the heat capacity at constant volume as 

\begin{equation}
C_V = \left(\frac{\partial U}{ \partial T}\right)\Big|_V 
\end{equation}

\noindent
where $U$ is the internal energy of the system, $U = H - PV$. After some straightforward algebra, it can be shown that

\begin{equation}
U = \frac{d-2}{4 G } \left( r_+^{d-3} - \frac{16 \pi G \sigma^2 r_+}{d-2} + \frac{d^2 - 6d + 7}{(d-2)^2} \psi_0 r^{d-2} \right),
\end{equation}

\noindent
which means that the internal energy does not depend of the pressure of the system. This is an expected result, since the internal energy is a function of the entropy and the volume, and both are functions only of the horizon radius. Thus we can conclude that $C_V$ vanishes, as in general relativity.

To study the heat engine for the Carnot and squared cycles presented in the previous section, it is not necessary to calculate the heat capacity at constant pressure, but we will calculate it for completeness. We cannot do in the same way as we did for general relativity, since we do not have a function of the form $T(S,P)$. However, for constant pressure, the heat involved in a process is given by the change of enthalpy of the process, i.e., we have that

\begin{equation}
C_P \equiv \frac{\delta Q}{d T}\Big|_P = \left(\frac{dH}{dr_+} \Big/ \frac{dT}{dr_+} \right)\Big|_{P(l)},
\end{equation}

\noindent
and it is given by 

\begin{widetext}
\begin{equation}
C_P^{f(R)} = C_P^{GR} + \psi_0 \frac{\kappa_b \Omega_{d-2}}{4 G \hbar} \tilde{S} \frac{(d^2-6d+7) \tilde{S}^{(d-3)/(d-2)} + (d-1)^2 \tilde{P} \tilde{S}^{(d-1)/(d-2)}}{(d-1) \tilde{S}\tilde{P} - (d-3) \tilde{S}^{(d-4)/(d-2)} + 16 \pi G \sigma^2 (d-3)/(d-2)} 
\end{equation}
\end{widetext}

One can now analyze the black hole as a heat engine in $f(R)$ gravity. We will perform the same analysis as in the case of general relativity, and to study how the efficiency of the heat engine is affected by the introduction of the string cloud, for different values of the parameter $\psi_0$. We will do this only for the squared cycle, since the temperature in $f(R)$ gravity is corrected only by a constant, thus maintaining the discussion at the same level, i.e., the efficiency of the Carnot cycle for $f(R)$ gravity will not be qualitatively different from the one in general relativity.

The study of the efficiency of the squared cycle in $f(R)$ gravity can be performed in several forms. As the theory has many free parameters, we will keep our discussion at four dimensions only, and maintain the same parameters we used in general relativity, namely,  $\tilde{S}_1 = \tilde{S}_4 = 1, \tilde{P}_1 = \tilde{P}_2 = 4, \tilde{S}_2 = \tilde{S}_3 = 3 $ and $\tilde{P}_3 = \tilde{P}_4 = 1$. With this choice, we are left with only two parameters one can modify, the string cloud parameter $\sigma$ and the $f(R)$ parameter $\psi_0$.

\begin{figure}[]
\centering
\begin{tabular}{@{}cc@{}}
\includegraphics[scale=0.40]{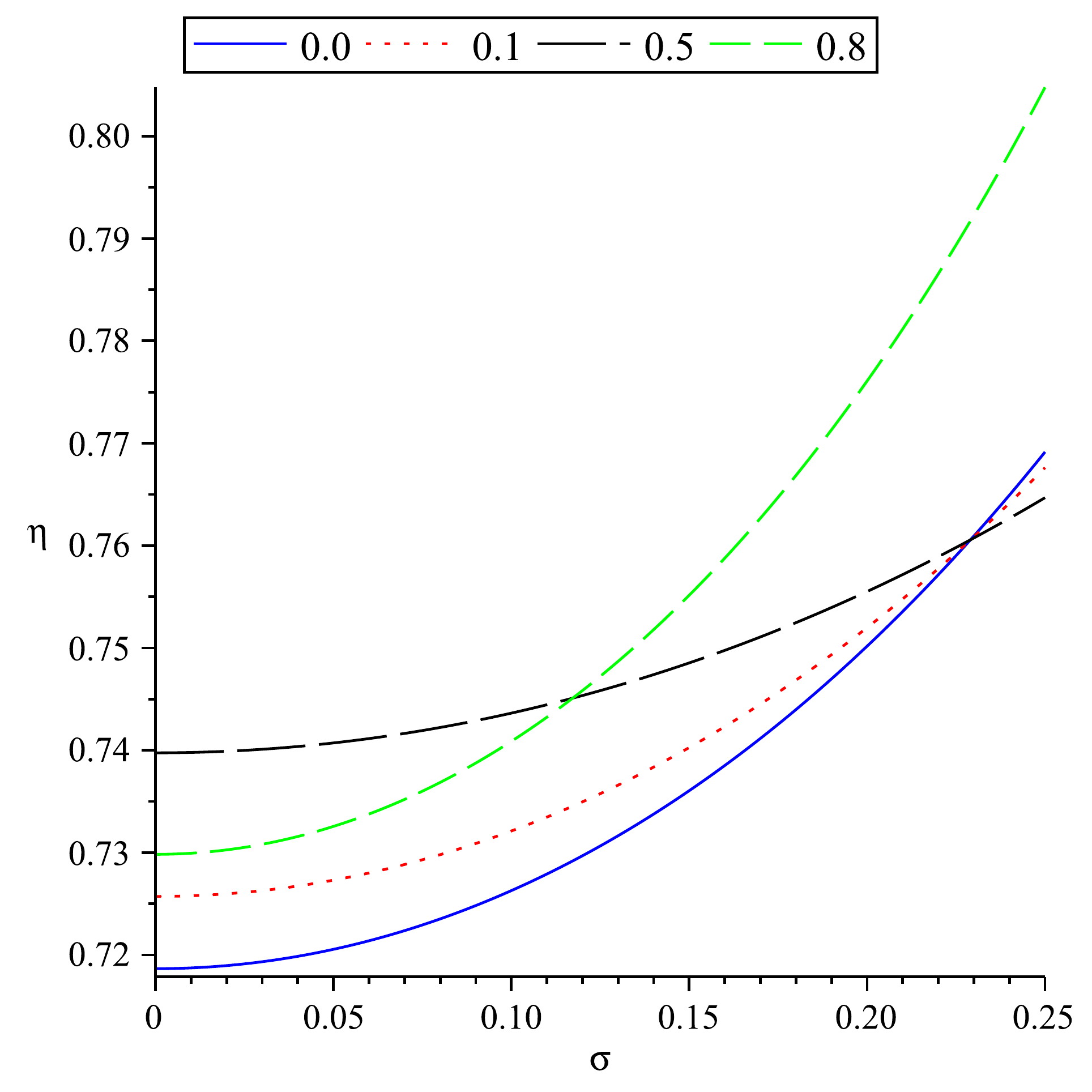}
\end{tabular}
\caption{Efficiency of the squared cycle as a function of the string cloud parameter $\sigma$ for $f(R)$ gravity, for several values of the parameter $\psi_0$.}
\label{fig:effSquaredFR1}
\end{figure}

\begin{figure}[]
\centering
\begin{tabular}{@{}cc@{}}
\includegraphics[scale=0.40]{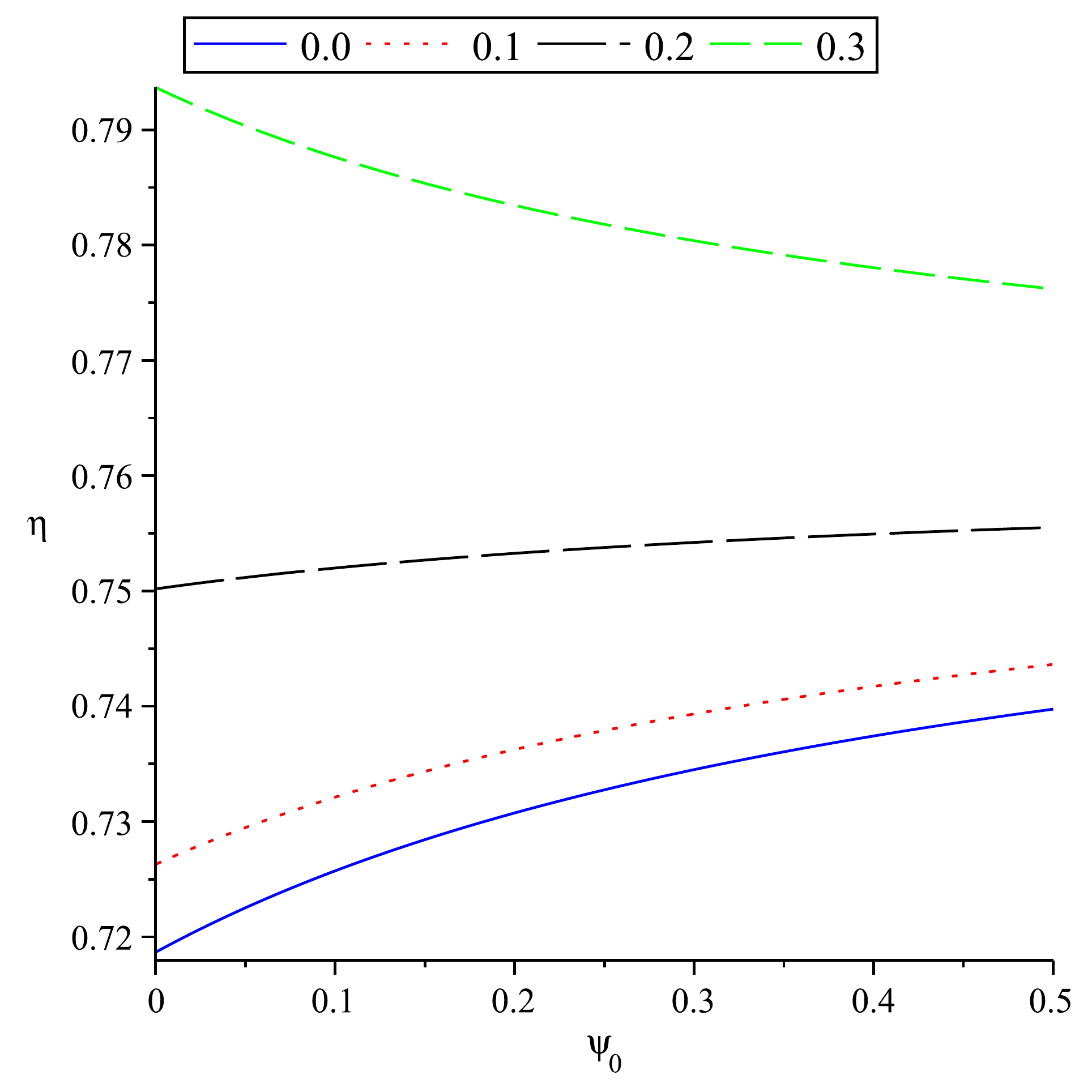}
\end{tabular}
\caption{Efficiency of the squared cycle as a function of the parameter $\psi_0$, related to $f(R)$ gravity, for several values of the parameter $\sigma$.}
\label{fig:effSquaredFR2}
\end{figure}

In figure (\ref{fig:effSquaredFR1}), we plot the efficiency of the heat engine as a function of the cloud parameter $\sigma$, for four different values of the parameter $\psi_0$, namely, $\psi_0 = 0, 0.1, 0.5$ and $0.8$. As one can see, the efficiency of the heat engine always increase, but not in a trivial manner. To reach a maximum efficiency, one has to choose a proper combination of the parameters $\sigma$ and $\psi_0$. In figure (\ref{fig:effSquaredFR2}) we show this relation in a different manner. This time, we vary the $f(R)$ parameter $\psi_0$ for four choices of the parameter $\sigma$, namely, $\sigma = 0.0, 0.1, 0.2$ and $0.3$. As one can see, the efficiency increases as the parameter $\sigma$ increases, but it can increase or decrease as the parameter $\psi_0$ increases. Once again this plot shows that, to find a maximum efficiency for the heat engine, one must find a proper combination of the parameters $\sigma$ and $\psi_0$. This set of parameters can be found, in principle, minimizing the analytic formula for the efficiency as a function of two parameters, but we are interested here only in the qualitative behavior of the theory, and on how the introduction of an $f(R)$ additional term affects the efficiency.

\section{Conclusions}

In this paper, we study the thermodynamics on the extended phase space of a black hole surrounded by a string cloud, both in general relativity and in the framework of $f(R)$ gravity. Our focus was to probe the $P-V$ criticality of the theory and the efficiency of these black holes as heat engines. For both scenarios, we have been able to find the thermodynamic equation of state related to the gravitational system, and to verify if the system presents a phase transition.

For general relativity, we obtained an explicit formula for the equations of state $P(V,T)$ and $T(S,P)$, and we were able to explicitly show that the heat capacity at constant volume is null. This means that one can consider isochorics as isobarics, a well-known feature for several others black holes. In the framework of $f(R)$ gravity, we have been unable to find an explicit equation of state of the form $T(S,P)$, but we conjecture that $C_V$ will also vanishes, since the internal energy depends only on the horizon radius, and not on the pressure. For the heat capacity at constant pressure, we have have found an explicit formula for both general relativity and $f(R)$ gravity.

An interesting result we obtained in the context of general relativity is that the relation between the critical pressure, volume and temperature is the same of the Van der Waals fluid, as long as the dimension of the spacetime is six. For others dimensions, we obtained an analytic formula for such relation. It is important to notice that such formula does not involve the string cloud parameter $\sigma$. This feature does not happens for $f(R)$ gravity, where in the relation there is an explicit dependence on the parameter $\sigma$.

Another interesting result is that we have found, for $f(R)$ gravity, a non-trivial formula for the thermodynamic volume of the theory, i.e., the conjugate quantity to the pressure. As long as we know, this is the first time a non-trivial thermodynamic volume is found for $f(R)$ gravity. The existence of non-trivial thermodynamic volumes is well-know in the literature, even for stationary black holes in general relativity, such as the Kerr-Newman spacetime. The meaning of such volumes is not clear until this moment.

Finally, we study the black  holes as heat engines and analyzed its efficiency, both for general relativity and for $f(R)$ gravity. We considered two cycles, the Carnot cycle and a squared cycle in the $P-V$ plane. The reason to use a squared cycle is that one can find an analytic formula for it, and exploit its properties graphically. We found that the efficiency of the Carnot cycle decreases as we introduce the string cloud, but it is not possible to conjecture that this is an universal behavior, since the temperature has a non trivial dependence on a large set of parameters. This argument is more clear when we move to $f(R)$ gravity, since it introduces a new parameter which can be varied. For the squared cycle, we showed that the efficiency increases as we increase the parameter $\sigma$, but the maximum efficiency appears to depend on a combination of both parameters, the one related to the string cloud and the one related to the gravitational theory. In any case, it is important to note that the introduction of a string cloud background will affect the behavior of the black hole as a heat engine. 
\section*{Acknowledgements}
JPMG and IPL thank Coordena\c c\~ao de Aperfei\c coamento de Pessoal de N\'ivel Superior (CAPES-Brazil), VBB thanks Conselho Nacional de Desenvolvimento Cient\'ifico e Tecnol\'ogico (CNPq-Brazil) and the work of H. Moradpour has been financially supported by Research Institute for Astronomy \& Astrophysics of Maragha (RIAAM) under research project No. $1/5440-58$.

\end{document}